\documentclass[aps,prl,showpacs,notitlepage,reprint,floatfix]{revtex4-1}
\usepackage{graphicx}
\usepackage{dcolumn}
\usepackage{amsmath}
\usepackage{bm}
\usepackage{todonotes}
\usepackage[utf8]{inputenc}
\usepackage{amsmath}
\usepackage{natbib}
\usepackage{graphicx}
\usepackage{caption}
\usepackage{subcaption}
\usepackage{lipsum}
\usepackage{float}

\newcommand{\bn}[1]{\mbox{\boldmath$#1$}}

\newcommand{\beq}{\begin{equation}}
\newcommand{\eeq}{\end {equation}}
\newcommand{\bea}{\begin{eqnarray}}
\newcommand{\eea}{\end{eqnarray}}

\begin{document}
\title{Intrinsic angular momentum, spin and helicity of higher-order Poincar\'e modes}
\author{{\rm {M. Babiker}}$^{1,*}$, K. Koksal$^{2}$, V. E. Lembessis$^{3}$ and J. Yuan$^{1}$}
\affiliation{$^1$School of Physics Engineering and Technology, University of York, YO10 5DD, UK}
\affiliation{$^2$Physics Department, Bitlis Eren University, Bitlis, Turkey}
\affiliation{$^3$Quantum Technology Group, Department of Physics and Astronomy, College of Science, King Saud University, Riyadh 11451, Saudi Arabia}
\vspace{10mm}
\footnote{$^*$ Corresponding author: m.babiker@york.ac.uk}
\vspace{10mm}
\vspace{10mm} 
\date{\today}

\begin{abstract}

The availability of coherent sources of higher order Poincar\'e optical beams have opened up new opportunities for applications such as in the optical trapping of atoms and small particles, the manipulation of chirally-sensitive systems and in improved encoding schemes for broad-bandwidth communications.  
Here we determine the intrinsic properties of Poincar\'e Laguerre-Gaussian (LG) modes which have so far neither been evaluated, nor their significance highlighted. The theoretical framework we adopt here is both novel and essential because it emphasises the crucial role played by the normally ignored axial components of the twisted light fields of these modes.  We show that the inclusion of the axial field components enables the intrinsic properties of the Poincar\'e  modes, notably their angular momentum, both spin and orbital as well as their helicity and chirality, to be determined. We predict significant enhancements of the intrinsic properties of these modes when compared with those due to the zero order LG modes. In particular, we show that higher order LG Poincar\'e modes exhibit super-chirality and, significantly so,  even in the case of the first order
 
\end{abstract}
\maketitle
\section{1. Introduction}

Recent advances in the field of twisted light have recognised the importance and the prospects for useful applications of the so-called higher order Poincar{\'e} vortex modes \cite{vaziri2002,milione2011,GALVEZ202195,maurer2007,Zhan2009,liu2014,fickler2014,krenn2014,ren2015,naidoo2016,liu2017,volpe2004,Holmes_2019,chen2020}.  For a given order specified by the  integer $m\ge 0$, the optical polarisation ${\bn {\hat \epsilon}}_m$ arises as a  non-separable superposition of the circular polarisation states $({\bn {\hat x}}\pm i{\bn {\hat y}})/\sqrt{2}$
with spatial phase $e^{\pm im\phi}$. Every state of polarisation is identified uniquely by a point $(\Theta_P,\Phi_P)$ on the surface of a unit higher order Poincar{\'e} sphere (HOPS), as shown and explained in the caption of Fig.1. 

Note that the positive order parameter $m$ includes the lowest orders, namely $m=0$ and $m=1$, to be referred to, respectively, as the basic 0th-order and first order set of modes.  Each is distinguished uniquely by its own order Poincar\'e sphere.  The Poincar\'e sphere for $m=0$ is scalar and incorporates all elliptically-polarised (including circularly- and linearly- polarised)  Gaussian modes.
The first order $m=1$ has a different Poincar\'e sphere and includes, in addition, the radially- and azimuthally-polarised optical vortex modes \cite{volpe2004}. 

Although several experimental reports have already confirmed the controlled generation of the higher order vector modes  \cite{naidoo2016,liu2017,chen2020}, the optical properties for arbitrary higher order $m\geq 0$ are, as far as we know, not yet been evaluated. Both the spin angular momentum (SAM) and the optical angular momentum (AM), as well as the orbital angular momentum (OAM) are yet to be evaluated for a general order $m$ Laguerre-Gaussian higher order modes.  \textcolor{black}{It is not clear what spatial density variations the vector components of the  SAM, OAM and AM of these modes have and what $m$-dependence each has.} 

It is also clearly  important to find out whether and how the higher order modes can lead to enhanced optical properties and enable new applications such as higher order encoding protocols for significantly increased bandwidth in quantum communication \cite{al2021structured}  and whether their angular momentum, spin and chirality are greatly enhanced. Evidently, enhanced optical chirality is highly desirable for chiroptical processes \cite{Tang2010}. We show that higher order Poincar\'e modes exhibit super-chirality and, significantly so,  even for the first order for which $m=1$.  
For recent accounts on the optical interaction with chiral matter the reader is referred to \cite{mun2020}.

The theoretical framework we adopt here aims first at evaluating the energy flux and the linear momentum density and proceeds to determine the cycle-averaged spin angular momentum density, the total angular momentum density, the orbital angular momentum density as well as the helicity and the chirality densities. For each of these main intrinsic properties we determine the spatial density distributions followed by the evaluation of its spatial integral over the focal plane at $z=0$.  The evaluations are carried out using the electromagnetic fields associated with the most general paraxial mode of arbitrary order $m$\textcolor{black}{($\geq 0 $)}.  An important feature of this paper is that although our main focus is on Laguerre-Gaussian modes our treatment is applicable to any type of twisted light and includes all the possible scenarios of optical polarisation of higher order Poincar\'e modes.

This paper is organised as follows.  In section 2 we begin with the vector potential ${\bf A}$ for a general paraxial optical vortex endowed with the higher order polarisation state ${\bn {\hat \epsilon}}_m$ which is defined in terms of the higher order Poincar\'e sphere.  We then describe the steps leading to the electric and magnetic fields of the higher order modes.
In section 3 we define the cycle-averaged intrinsic properties and proceed to evaluate their respective densities for the higher order modes for each of the stated properties, namely the SAM, the AM and the OAM, as well as the helicity and the chirality. Each evaluated density is then followed by the evaluation of the integrated (total) property per unit length. The $m$-dependence of the helicity density has been elucidated in \cite{babiker24}, so the emphasis in this paper is the demonstration ofof the involvement of both the higher order parameter $m$ as well as the radial number $p$, as for Laguerre-Gaussian modes LG$^m_p$.  Section 4 summarises the results and  comments on the significance of these results and Appendices A and B  discuss the normalisation factor ${\cal A}_0$ in terms of the Power ${\cal P}_T$ and we consider the four definite integrals needed in the body of the paper.

\section{2. Fields of higher order Poincar\'e modes}
The paraxial electromagnetic fields associated with an optical vortex in a higher order state of polarisation are derivable from a vector potential in cylindrical coordinates ${\bf r}=(\rho,\phi,z)$ in the form
\beq
{\bf A}={\bn {\hat \epsilon}}_{m,p}{\tilde {\cal F}}_{m, p}(\rho,z)e^{ik_zz}
\label{vect1}
\eeq
where  ${\bn {\hat \epsilon}}_{m,p}$ is the order $m$ polarisation and its variations cover every point  on the surface of the order order $m\geq 0$ unit Poincar\'e sphere. This is  as follows
\begin{widetext}
\beq
{\bn {\hat \epsilon}}_{m,p}=e^{im\phi}\frac{({\bn {\hat x}}-i{\bn {\hat y}})}{\sqrt{2}}\cos\left(\frac{\Theta_P}{2}\right)e^{-i\Phi_P/2}+e^{-im\phi}\frac{({\bn {\hat x}}+i{\bn {\hat y}})}{\sqrt{2}}\sin{\left(\frac{\Theta_P}{2}\right)}e^{i\Phi_P/2}
\label{epsil}
\eeq
\end{widetext}
where $\Theta_P$ and $\Phi_P$ are Poincare sphere angles as defined in Figure 1. The vector ${\bn {\hat \epsilon}}_m$ is the most general polarisation state vector, and written using our convention as depicted in Fig. 1 so it differs slightly from the form defined by Milione et al \cite{milione2011}. The validity of the polarisation states for the Poincar\'e modes, has already been confirmed experimentally \cite{liu2014,naidoo2016,chen2020}. 

In Eq.(\ref{vect1}) the polarisation is seen multiplying the paraxial vortex mode function ${\tilde{\cal F}}_{m, p}(\rho,z)e^{ik_zz}$ which has no $\phi$-dependence and in which $k_z$ is the wavevector for the light travelling along the $+z$ axis and ${\cal F}_{m, p}$  specifying only the amplitude variations in terms of the coordinates $(\rho,z)$. The mode function  is  labelled by the its integer order $m>0$, (which is also the winding number) and  $p\geq 0$ (which is the radial number), as for a Laguerre-Gaussian (LG) optical vortex modes. 


\begin{widetext}
    
\begin{figure}
\includegraphics[width=0.8\columnwidth]{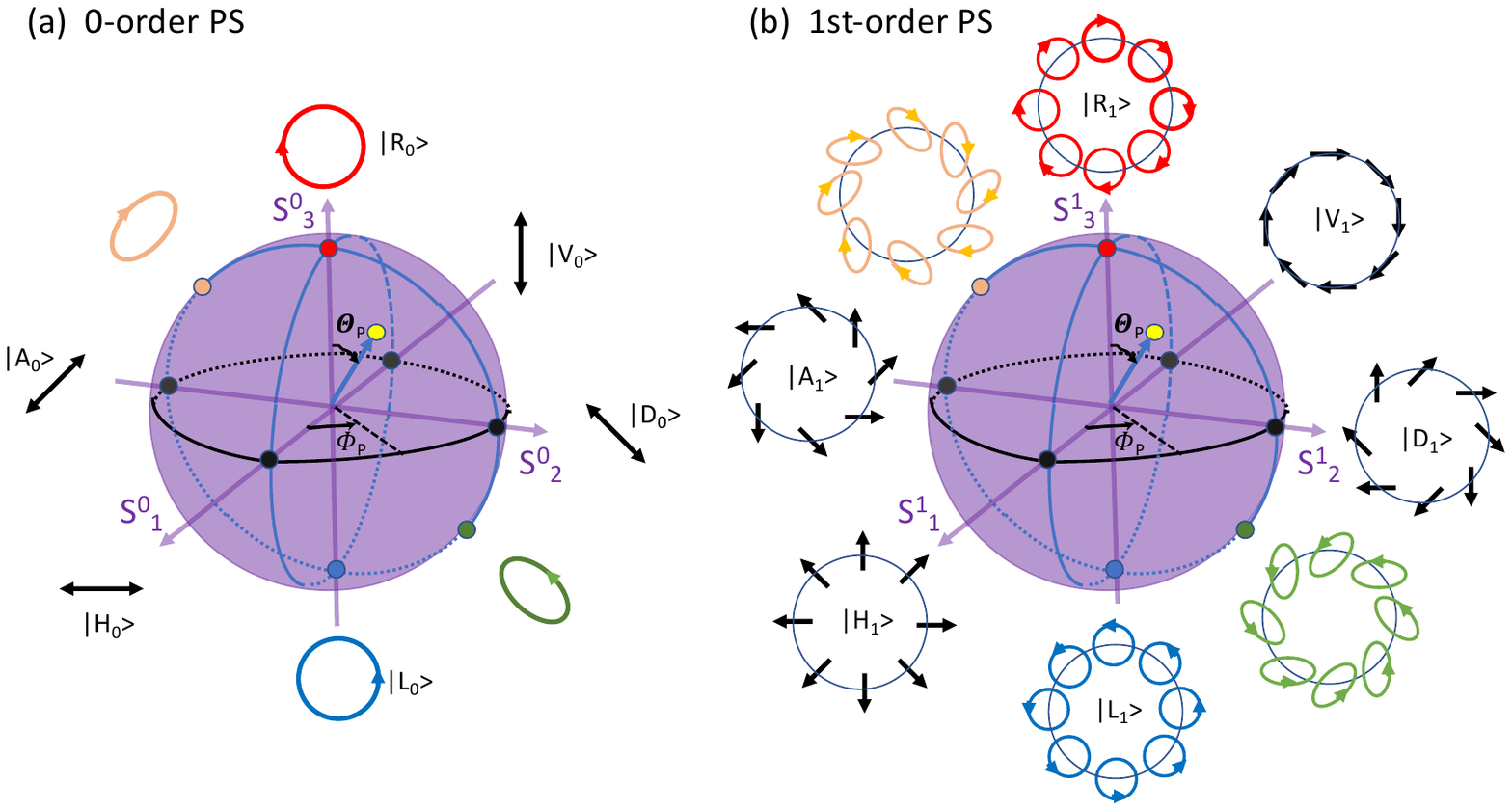}
\caption{0th order, (a), and 1st-order, (b), Pioncar\'e Sphere (PS) representation of the  polarisation state in which optical polarisation is coupled with vortex phase, characterized by a unit sphere with respect to the corresponding Stokes-parameter(-like) Cartesian coordinates ($S^0_1, S^0_2$, $S^0_3$) and ($S^1_1, S^1_2$, $S^1_3$) respectively.  It is seen that the 0th order PS is equivalent to the conventional PS where $|H>$ and $|V>$ are commonly used to denote the vertically and horizontally linearly polarized light, $|A>$ and $|D>$ for $\pm 45^o$ tilted linearly polarized light, $|R>$ and $|L>$ for right-hand and left-hand circularly-polarized light, respectively.   The 1st order PS figure is related to the corresponding figure by Milione et al \cite{milione2011} with slightly different conventions for $S^1_1$ and $S^1_2$.  Six sets of special vector modes are drawn in different colours next to each sphere for illustration.  Their positions on the Poincar\'e sphere are indicated by dots of the same colour.}
\end{figure}
\end{widetext}

In order to simplify the evaluations we now express the polarisation vector ${\bn {\hat \epsilon}}_m$, Eq.(\ref{epsil}), in the following form
\beq
{\bn {\hat \epsilon}}_{m,p}=e^{im\phi}({\bn {\hat x}}-i{\bn {\hat y}}){\cal U}_P+e^{-im\phi}({\bn {\hat x}}+i{\bn {\hat y}}){\cal V}_P\label{epsil}
\eeq
where ${\cal U}_P$ and ${\cal V}_P$ are the complex Poincare' angular functions (i.e. of $\Theta_P,\Phi_P$) given by
\beq 
 {\cal U}_P=\frac{1}{\sqrt{2}}\cos{\left(\frac{\Theta_P}{2}\right)}e^{i\Phi_P/2}
 ;\;\;\;\;{\cal V}_P=\frac{1}{\sqrt{2}}\sin{\left(\frac{\Theta_P}{2}\right)}e^{-i\Phi_P/2}
 \label{upee}
\eeq

The Poincare angular functions satisfy the following identities
 \bea  
 ({\cal U}_P^*{\cal V}_P-{\cal V}_P^*{\cal U}_P)&=&-\frac{i}{2}\sin(\Theta_P)\sin(\Phi_P)\nonumber\\
 (|{\cal U}_P|^2-|{\cal V}_P|^2)&=&\frac{1}{2}\cos(\Theta_P)\nonumber\\
 (|{\cal U}_P|^2+|{\cal V}_P|^2)&=&\frac{1}{2}
 \label{poincare}
 \eea

We seek to develop the analysis for a general ${\tilde{\cal F}}$, which could be appropriate for any optical vortex type, for example Laguerre-Gaussian, Bessel, Bessel-Gaussian,..etc., and then consider the evaluations for the Laguerre-Gaussian case.
We begin by writing the vector potential in Eq.(\ref{vect1}) in the following form
\begin{widetext}
\beq
{\bf A}=\left\{({\bn {\hat x}}-i{\bn {\hat y}})e^{im\phi}{\cal U}_P+({\bn {\hat x}}+i{\bn {\hat y}})e^{-im\phi}{\cal V}_P\right\}{\tilde {\cal F}}_{m,p}(\rho,z)e^{ik_zz}
\label{vect2}
\eeq
\end{widetext}
This form of the generally-polarised mode of order $m$ suggests that it is a  superposition of two vortex modes, one with winding number $m$ and right-handed circular polarisation, weighted by the Poincar\'e function ${\cal U}_P$  and the second has winding number $-m$ and left-handed circular polarisation, weighted by the Poincar\'e function ${\cal V}_P$.

One of the main requirements to be satisfied by free-space paraxial optical fields is that the electric field must be derivable from the magnetic field using the Maxwell curl equation and that the electric field must produce the same magnetic field via the second Maxwell curl equation.  \textcolor{black}{For convenience, we can do so by writing the vector potential Eq.(\ref{vect2}) as the sum of two parts as follows:}
\beq
{\bf A}={\bf A_1}+{\bf A_2}\label{combvec}
\eeq
where
\begin{widetext}
\beq
{\bf A_1}=({\bn {\hat x}}-i{\bn {\hat y}}){\cal F}^{(1)}_{m, p}({\bf r})e^{ik_zz};\;\;\;\;
{\bf A_2}=({\bn {\hat x}}+i{\bn {\hat y}}){\cal F}^{(2)}_{m, p}({\bf r})e^{ik_zz}\label{vect3}
\eeq
\beq
{\cal F}^{(1)}_{m,p}({\bf r})={\cal U}_P e^{im\phi}{\tilde {\cal F}}_{m, p}(\rho,z);\;\;\;
{\cal F}^{(2)}_{m,p}({\bf r})={\cal V}_P e^{-im\phi}{\tilde {\cal F}}_{m, p}(\rho,z)
\label{eff1,2}
\eeq
\end{widetext}

Note that so far we have regarded ${\tilde{\cal F}}$ as  a general optical vortex mode amplitude function and here we develop the formalism without specifying this function.  Once the results are arrived at, we shall consider the special case of ${\tilde{\cal F}}$ appropriate for Laguerre-Gaussian modes. 

The magnetic and electric  fields of our generally-polarised mode are written as the sums of two terms ${\bf B}={\bf B}_1+{\bf B}_2$ and ${\bf E}={\bf E}_1+{\bf E}_2$ where ${\bf B}_i={\bn {\nabla}}\times{\bf A}_i;\;\;\;i=1,2$. The sequence of steps leading to the required expressions for the fields involve dealing with the two parts of the magnetic field first and from those use Maxwell's curl B equation to derive the corresponding electric field parts.  We have 
\bea
{\bf B}_1&=&ik_z({\bn {\hat y}}+i{\bn {\hat x}}){\cal F}^{(1)} e^{ik_zz}-{\bn {\hat z}}\left(i\frac{\partial {\cal F}^{(1)}}{\partial x}+\frac{\partial {\cal F}^{(1)}}{\partial y}\right)e^{ik_zz}\nonumber\\
{\bf E}_1&=&ick_z({\bn {\hat x}}-i{\bn {\hat y}}){\cal F}^{(1)} e^{ik_zz}-{\bn {\hat z}}c\left\{\frac{\partial {\cal F}^{(1)}}{\partial x}-i\frac{\partial {\cal F}^{(1)}}{\partial y}\right\}e^{ik_zz}\nonumber\\
{\bf B}_2&=&ik_z({\bn {\hat y}}-i{\bn {\hat x}}){\cal F}^{(2)} e^{ik_zz}+{\bn {\hat z}}\left(i\frac{\partial {\cal F}^{(2)}}{\partial x}-\frac{\partial {\cal F}^{(2)}}{\partial y}\right)e^{ik_zz}\nonumber\\
{\bf E}_2&=&ick_z({\bn {\hat x}}+i{\bn {\hat y}}){\cal F}^ {(2)}e^{ik_zz}-{\bn {\hat z}}c\left\{\frac{\partial {\cal F}^{(2)}}{\partial x}+i\frac{\partial {\cal F}^{(2)}}{\partial y}\right\}e^{ik_zz}\nonumber\\
\label{emfields}
\eea
where we have dropped the subscript labels in ${\cal F}^{(1),(2)}$ and in ${\tilde {\cal F}}$ for ease of notation and these can be restored when the need arises. The fields we have derived in the set of equations (\ref{emfields}) form the basis for the derivation of the optical properties of the order $m$ vector vortex mode. Note that up to now we have not specified the type of vortex mode and we continue to deal with the general form involving the vector potential amplitude function ${\tilde{\cal F}}_{m,p}(\rho,z)$.
\\
\section{3. Cycle-averaged intrinsic properties}

Our aim is to evaluate all the major cycle-averaged properties of the higher order Poincar\'e mode. These include the optical spin angular momentum (SAM) density ${\bar {\bf s} }$ the angular momentum (AM) density ${\bn {\bar j}}$ from which we obtain the orbital angular momentum (OAM) density and finally, the helicity and the chirality of the mode. The cycle-averaged AM density requires evaluation of the linear momentum density and we also need to evaluate the power flux for normalisation.  The relevant properties are defined as follows \cite{allen1999}
\bea
{\bn {\bar s}}&=&\frac{1}{4\omega}\Im\left\{[\epsilon_0{\bf E}^*\times{\bf E}]+\frac{1}{\mu_0}[{\bf B}^*\times{\bf B}]\right\};\nonumber\\
&=&{\bn {\bar s}}_{E}+{\bn {\bar s}}_{B}\;\;\;\;\;\;{\rm {(SAM \;density)}}
\label{sama}
\eea
\beq
{\bn {\bar j}}={\bf r}\times {\bn {\bar {\pi}}};\;\;\;\;({\rm {AM\; density}})\label{jbar}
\eeq
\beq 
{\bn {\bar \ell}}={\bn {\bar j}}-{\bn {\bar s}};\;\;\;\;\;({\rm {OAM\; density}})\label{ellbar}
\eeq

\bea
{\bar {\eta}}({\bf r})&=&\frac{c}{\omega^2}{\bar \chi}=-\frac{\epsilon_0 c}{2\omega}\Im{[{\bf E}^*\cdot{\bf B}]}\nonumber\\
&&({\rm {Helicity/\;Chirality\; density}}),
\label{hel}
\eea
where in the above ${\bar {\bn {\pi}}}=\frac{1}{c^2}{\bar {\bf w}}$ is the linear momentum density with ${\bar {\bf w}}= \frac{1}{2\mu_0}\Re[{\bf E}^*\times{\bf B}]$ the energy density. 
The symbols ${\Re}[...]$ and $\Im[...]$ stand for real and imaginary parts of [...] and the superscript * in ${\bf E}^*$ stands for the complex conjugate of ${\bf E}$. Note that in free space the chirality density ${\bar \chi}$ is proportional to the helicity density ${\bar {\eta}}({\bf r})$, so we only need to deal with the helicity from which the chirality follows.

As stated above we consider the evaluations of the above densities and deal with them in turn, including their vector components, all evaluated specifically in relation to the higher order Poincar\'e modes. After this the next tasks will involve the evaluation of the total intrinsic properties per unit length  for SAM and AM and from these two we deduce the OAM.  Finally we evaluate the helicity and chirality per unit length. Each property is evaluated as the space integral of the corresponding density over the x-y plane.

The evaluation now requires as a first step expressions for the  $x$- and $y$-derivatives of ${\cal F}^{(1)}$ in polar coordinates.  Note that ${\cal F}^{(1)}$ is distinguished by the phase factor $\exp{(+im\phi)}$ and ${\cal F}^{(2)}$ is distinguished by the phase factor $\exp{(-im\phi)}$. We have for ${\cal F}^{(1)}$
\beq 
\frac{\partial {\cal F}^{(1)}}{\partial x}={\cal U}_P\left\{\cos\phi{\tilde {\cal F}'}-i\frac{m}{\rho}\sin\phi {\tilde {\cal F}}\right\}e^{im\phi}\label{dfx}
\eeq
\beq 
\frac{\partial {\cal F}^{(1)}}{\partial y}={\cal U}_P\left\{\sin\phi{\tilde {\cal F}'}+i\frac{m}{\rho}\cos\phi {\tilde {\cal F}}\right\}e^{im\phi}\label{dfy}
\eeq
\beq 
\frac{\partial {\cal F}^{(2)}}{\partial x}={\cal V}_P\left\{\cos\phi{\tilde {\cal F}'}+i\frac{m}{\rho}\sin\phi {\tilde {\cal F}}\right\}e^{-im\phi}\label{dfx2}
\eeq
and 
\beq 
\frac{\partial {\cal F}^{(2)}}{\partial y}={\cal V}_P\left\{\sin\phi{\tilde {\cal F}'}-i\frac{m}{\rho}\cos\phi {\tilde {\cal F}}\right\}e^{-im\phi}\label{dfy2}
\eeq

\subsection{Paraxial Laguerre-Gaussian forms}

Ultimately we will apply the formalism to the case of higher order Laguerre-Gaussian Poincar\'e modes.  We thus require the amplitude function of these for  winding number $m$ and radial number $p$ which is
\begin{widetext}
\beq
{\tilde {\cal F}}_{m,p}(\rho,z)={\cal A}_0\frac{C_{|m|,p}}{\sqrt{1+z^{2}/z_{R}^{2}}}\left( \frac{\rho \sqrt{2}}{w_{0} \sqrt{1+z^{2}/z_{R}^{2}}}\right)^{|m|}
\exp\left[\frac{-\rho^{2}}{w_{0}^{2}(1+z^{2}/z_{R}^{2})}\right]L_{p}^{|m|}\left\{\frac{2\rho^{2}}{w_{0}^{2}(1+z^{2}/z_{R}^{2})}\right\}e^{i\xi(\rho, z)},
\label{eq:refname3}
\eeq
\end{widetext}
where ${\cal A}_0$ is a normalisation factor, to be determined in terms of the power ${\cal P}_T$. The phase function $\xi(\rho, z)$  includes the Gouy and the curvature phases 
\beq
\xi(\rho, z)=- (2p+|m|+1)\arctan\left(\frac{z}{z_{R}}\right)+\frac{kz\rho^{2}}{2(z^2+z_{R}^{2})}.
\label{eq:refname4}
\end{equation}
Here $w_{0}$ is the beam waist, $z_{R}=w_0^2k/2$ is the Rayleigh range, $C_{|m|,p}=\sqrt{p!/(p+|m|)!}$ and $L_{p}^{|m|}$ the associated Laguerre polynomial.

\textcolor{black}{Note that the amplitude function ${\tilde{\cal F}}$ for LG modes only depends on the absolute value of $m$, so that we can now interpret the $m$-th order Poincar\'e modes, defined in in  (\ref{combvec}, \ref{vect3}, \ref{eff1,2}), as LG$_{m,p}$ and LG$_{-m,p}$ modes.  This fact is useful experimentally as well as in aiding the interpretation of our results.}

Our main concern will be  on evaluating the variations of each intrinsic property on the focal plane at $z=0$ so that the vector potential amplitude function reduces to 
\beq
{\tilde {\cal F}}_{m,p}(\rho)={\cal A}_0\sqrt{\frac{p!}{(p+|m|)!}} e^{-\frac{\rho^2}{w_0^2}}  
\left(\frac{\sqrt{2}\rho}{w_0}\right)^{|m| }L^{|m|}_p\left(\frac{2\rho^2}{w_0^2}\right),\label{laguerre}
\eeq
where now ${\tilde {\cal F}}_{m,p}$ depends only on the radial coordinate $\rho$. In Appendix A we evaluate the normalisation constant ${\cal A}_0$ and express it in terms of the applied power ${\cal P}_T$.

\subsection{The SAM density}

The SAM density Eq.(\ref{sama}) receives contributions from both the electric field ${\bar {\bf s}}_{E}$ and the magnetic field ${\bar {\bf s}}_B$. We  begin by the evaluation of the first part ${\bar {\bf s}}_E$ of the  SAM density which satisfies
\beq
\frac{4\omega}{\epsilon_0}{\bar {\bf s}}_E=\Im[{\bf E}^*\times{\bf E}].
\eeq
Since ${\bf E}={\bf E}_1+{\bf E}_2$, the vector cross product contains four products, two of which are direct terms and two mixed terms 
\beq
{\bf E}^*\times{\bf E}=\left\{({\bf E}_1^*\times{\bf E}_1+{\bf E}_2^*\times{\bf E}_2)+({\bf E}_1^*\times{\bf E}_2+{\bf E}_2^*\times{\bf E}_1)\right\},
\eeq
where the first set of terms are direct and the second set are the mixed terms.
A similar treatment is followed for the second part of the SAM density. We have
\beq
4\omega\mu_0{\bar {\bf s}}_B=\Im[{\bf B}^*\times{\bf B}].
\eeq
Here also the vector product consists of four products two of which are direct and two are mixed
\beq
{\bf B}^*\times{\bf B}=({\bf B}_1^*\times{\bf B}_1+{\bf B}_2^*\times{\bf B}_2)+({\bf B}_1^*\times{\bf B}_2+{\bf B}_2^*\times{\bf B}_1).
\eeq

The evaluation of the SAM density is straightforward, but rather lengthy.  Each of the eight vector products consists of three vector components which we evaluated separately by direct substitutions of the electric and magnetic fields given in Eqs.(\ref{emfields}). 

Extensive careful evaluations lead to the following results. We have found that the direct E terms are equal to the direct B terms
\beq   
\left[{\bf E}_{1,2}^*\times{\bf E}_{1,2}\right]=c^2\left[{\bf B}_{1,2}^*\times{\bf B}_{1,2}\right]\label{equal}.
\eeq
We have also found that all components of the mixed E terms are equal and opposite in sign to those of the B term.  We have 
\beq   
\left[{\bf E}_1^*\times{\bf E}_2\right]=-c^2\left[{\bf B}_1^*\times{\bf B}_2\right]\label{noequal1}.
\eeq
Similarly we have found
\beq   
\left[{\bf E}_2^*\times{\bf E}_1\right]=-c^2\left[{\bf B}_2^*\times{\bf B}_1\right]\label{noequal2}.
\eeq
Thus the direct terms add while mixed ones cancel. Making use of the above relations we get the simple result
\beq   
[{\bf E}^*\times{\bf E}]+c^2[{\bf B}^*\times{\bf B}]=2\left\{[{\bf E}_1^*\times{\bf E}_1]+[{\bf E}_2^*\times{\bf E}_2]\right\}.
\eeq
This simplifies the analysis.


\subsection{Transverse SAM density components}
The x-component of the spin density is 
\beq
{\bar s}_x=\frac{\epsilon_0}{2\omega}\Im\left\{[{\bf E}_1^*\times{\bf E}_1]_x+[{\bf E}_2^*\times{\bf E}_2]_x\right\}.
\eeq

Direct evaluations using the expressions in Eq.(\ref{emfields}) for ${\bf E}_1$ and ${\bf E}_2$ lead us to the result
\beq
\textcolor{black}{{\bar s}_x=\frac{\epsilon_0c^2k_z}{2\omega}\left\{m\frac{{\tilde {\cal F}}^2}{\rho}+{\tilde {\cal F}'}{\tilde {\cal F}}\right\}\sin\phi},
\eeq
where we have made use of the identities in Eqs.(\ref{poincare}). Similarly on evaluating the y-component and obtain
\beq
\textcolor{black}{{\bar s}_y=-\frac{\epsilon_0c^2k_z}{2\omega}\left\{m\frac{{\tilde {\cal F}}^2}{\rho}+{\tilde {\cal F}'}{\tilde {\cal F}}\right\}\cos\phi,}
\eeq

These combine to give the transverse SAM vector component ${\bn {\bar s}}_{\bot}$ defined as follows
\beq
{\bn {\bar s}}_{\bot}={\bar s}_x{\bn {\hat x}}+{\bar s}_y{\bn {\hat y}}.
\eeq

This result can be written in terms of the azimuthal unit vector ${\bn {\hat \phi}}$ as follows
\beq  
\textcolor{black}{{\bn {\bar s}}_{\bot}
=-\frac{\epsilon_0c^2k_z}{2\omega}
\left\{m\frac{{\tilde {\cal F}}^2}{\rho}+{\tilde {\cal F}'}{\tilde {\cal F}}\right\}{\bn {\hat{\phi}}}
\label{transverse}}
\eeq
\textcolor{black}{The emergence of the transverse spin is  a direct consequence of the longitudinal field components in Eqs. (\ref{emfields})  whose strength involves both the $\phi$ and $\rho$ gradients in the transverse plane \cite{Aiello2015}. The $\phi -$gradient accounts for the $m-$dependence in the first term, while the radial gradient is manifest in the second term ${\hat{\cal F}'}{\hat{\cal F}}$.} 
\\
\subsection{The longitudinal component of the SAM density}
The longitudinal spin density component is given by
\beq  
{\bar s}_z=-\frac{\epsilon_0}{2\omega}\Im\left\{[{\bf E}_1^*\times{\bf E}_1]_z+[{\bf E}_2^*\times{\bf E}_2]_z\right\}.
\label{sbarzed}
\eeq
Similar evaluations lead to the result 
\bea  
{\bar s}_z&=&\frac{\epsilon_0c^2k_z}{2\omega}[|{\cal U}_P^2|-|{\cal V}_P^2|]{\tilde {\cal F}}^2\nonumber\\
&=&\frac{\epsilon_0c^2k_z}{2\omega}\cos(\Theta_P){\tilde {\cal F}}^2.\label{sbarrz}
\label{longit}
\eea
\textcolor{black}{The lack of any $m-$dependence in ${\bar s}_z$ in Eq. (\ref{sbarrz}) is easy to see since ${\bar s}_z$ depends only on contributions arising from the transverse field components.}

\begin{figure}
\includegraphics[width=\columnwidth]{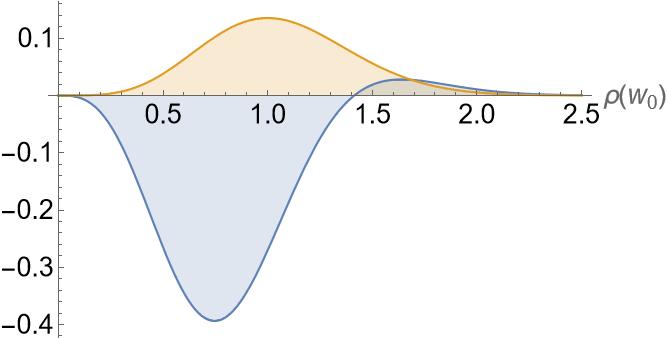}
\caption{The variations of the SAM density components (arbitrary units) with the radial coordinate $\rho$ (in units of $w_0)$.  Here the blue curve represents the magnitude of the transverse component (Eq. \ref{transverse})  and the yellow curve represents the longitudinal $z$ component (Eq. \ref{longit}).  For illustration we have chosen to consider the order $m=2$ and the point on the north pole $\Theta_p=0,\Phi_p=0$ of the order m=2 Poincar\'e sphere.}
\label{fig:saml2m3}
\end{figure}

This completes the evaluations of the general form of the transverse and the longitudinal components of the SAM density, which are applicable to any optical vortex amplitude function ${\tilde {\cal F}}$. However, we shall proceed to consider the specific case in which ${\tilde {\cal F}}$ corresponds to a Laguerre-Gaussian amplitude function. 
The variations of the SAM density components with the radial coordinates are displayed in Fig.(\ref{fig:saml2m3}).

\subsection{Integrated SAM density}
We have seen that the transverse SAM density components are proportional to sine and cosine of the azimuthal angle $\phi$  and so on spatial integration they vanish
\beq
{\bar S}_x=0;\;\;\;\;{\bar S}_y=0.
\eeq
Only the z-component survives on integration, so we have, on substituting from Eq.(\ref{sbarzed})
\beq 
{\bar S}_z=\int_0^{2\pi}d\phi\int_0^{\infty}{\bar s}_z\rho\;d\rho= \frac{\pi\epsilon_0}{\omega}\cos(\Theta_P){\cal I}_P,
\eeq
and on substituting for ${\cal I}_P$ in terms of ${\cal P}_T$ using Eq.(\ref{peeT}) we have
\beq 
{\bar S}_z=\frac{c^2k_z^2\pi\epsilon_0}{\omega}\cos(\Theta_P)\frac{\mu_0{\cal P}_T}{\pi c k_z^2},
\eeq
which finally leads us to the total SAM per unit length in the form 
\beq  
{\bar S}_z=\cos(\Theta_P)\left(\frac{{\cal P}_T}{\omega k_z}\right).
\eeq
The expression between the brackets has the dimensions of angular momentum per unit length. Note that this is independent of  $m$, which means that all higher order modes have the invariant SAM which depends only on the Poincare angle $\Theta_P$. The factor $\cos(\Theta_P)$ takes the values $\pm 1$ at $\Theta_P=0,\pi$ as for circularly-polarised modes and is zero for $\Theta_P=\pi/2$, as for radially and azimuthally-polarised modes.  Other values of $\Theta_P$ concern elliptically-polarised modes.\\

\subsection{AM desnity}
The cycle-averaged angular momentum density is defined as follows:
\beq
{\bn {\bar j}}={\bf r}\times {\bar {\bn {\pi}}},
\eeq
where ${\bar {\bn {\pi}}}=\frac{1}{c^2}{\bar {\bf w}}$ is the linear momentum density with ${\bar {\bf w}}= \frac{1}{2\mu_0}\Re[{\bf E}^*\times{\bf B}]$ the energy density. 
Thus in order to proceed we have to work out all the components of the Poynting vector. The vector product $[{\bf E}^*\times{\bf B}]$ where the fields are ${\bf E}={\bf E}_1+{\bf E}_2$ and ${\bf B}={\bf B}_1+{\bf B}_2$ is
\beq
[{\bf E}^*\times{\bf B}]=[{\bf E}_{1}^*\times{\bf B}_{1}]+[{\bf E}_2^*\times{\bf B}_2]+[{\bf E}_1^*\times{\bf B}_2]+[{\bf E}_2^*\times{\bf B}_1].
\label{edotb2}
\eeq
We refer to the components of the different terms generically by ${\cal P}_{\alpha\beta i}$ where $\alpha,\beta$ each take the values 1 and 2 and $i=(x,y,z)$. 

As an example to how evaluations proceed, we consider ${\cal P}_{11x}$.  We have
\beq  
{\cal P}_{11x}=[{\bf E}_{1}^*\times{\bf B}_{1}]_x=E_{1y}^*B_{1z}-E_{1z}^*B_{1y},
\eeq
which gives
\begin{widetext}
\bea  
{\cal P}_{11x}&=&(-ck_z)\left\{{{\cal F}^{(1)}}^*\left[i\left(\frac{\partial{\cal F}^{(1)}}{\partial x}\right)+\left(\frac{\partial{\cal F}^{(1)}}{\partial y} \right)\right]+{\cal F}^{(1)}\left[-i\left(\frac{\partial{\cal F}^{(1)}}{\partial x}\right)^*+\left(\frac{\partial{\cal F}^{(1)}}{\partial y}\right)^*\right]\right\}.
\eea
\end{widetext}
We obtain on substituting for the derivatives 
\beq  
{\cal P}_{11x}=-2ck_z|{\cal U}_P|^2\left({\tilde {\cal F}}'{\tilde {\cal F}}+\frac{m}{\rho}{\tilde {\cal F}}^2\right)\sin\phi.
\eeq
Similar evaluations were carried out for all 16 terms in Eq.(\ref{edotb2}). It turns out that all the 8 mixed term components arising from $[{\bf E}_1^*\times{\bf B}_2]+[{\bf E}_2^*\times{\bf B}_1]$ are imaginary and so will not contribute to the real part of of the overall Poynting vector.   From these we are able to  evaluate the components of ${\bar {\bf j}}$.  For example the y-component ${\bar j}_y$ is given by
\beq
{\bar j}_y=\frac{\epsilon_0 c^2 k_z}{2\omega}\left\{z({\cal P}_{11x}+{\cal P}_{22x})-x({\cal P}_{11z}+{\cal P}_{22z})\right\}.
\eeq
The x-component ${\bar j}_x$  together with the y-component ${\bar j}_y$, constitute the transverse angular momentum density vector along ${\bn {\hat {\phi}}}$
\beq
\textcolor{black}{\bn {{\bar {j}}}_{\bot}=\frac{\epsilon_0 c^2k_z^3}{2 \omega}\rho {\tilde {\cal F}}^2}{\bn {\hat {\phi}}}\label{jayphi}
\eeq
Similar evaluations of the longitudinal  (z-component) of the angular momentum density ${\bar j}_z$ in the form 
\beq
{\bar j}_z=\frac{\epsilon_0 c k_z}{2\omega}\left\{x({\cal P}_{11y}+{\cal P}_{22y})-y({\cal P}_{11x}+{\cal P}_{22x})\right\}.
\label{jayzee0}
\eeq
We obtain
\beq
\textcolor{black}{{\bar j}_z=\frac{\epsilon_0c^2k_z^2}{2\omega}\left\{m{\tilde {\cal F}}^2+\rho{\tilde {\cal F}}{\tilde {\cal F}}'\right\}\cos\Theta_P}.\label{jayzee}
\eeq

 The angular momentum density vector is therefore given by
 \beq
 {\bn {\bar j}}={{\bar j}}_{\bot}{\bn {\hat \phi}}+{\bar j}_z{\bn{\hat z}},
 \eeq
 where ${{\bar j}}_{\bot}$ is the magnitude of the transverse component, given by Eq.(\ref{jayphi}), and ${\bar j}_z$ is the longitudinal component, given by Eq.(\ref{jayzee}).
 
 \subsection{Variations of ${{\bar j}}_{\bot}$ and ${\bar j}_z$ with $\rho$}

\begin{figure}
\includegraphics[width=\columnwidth]{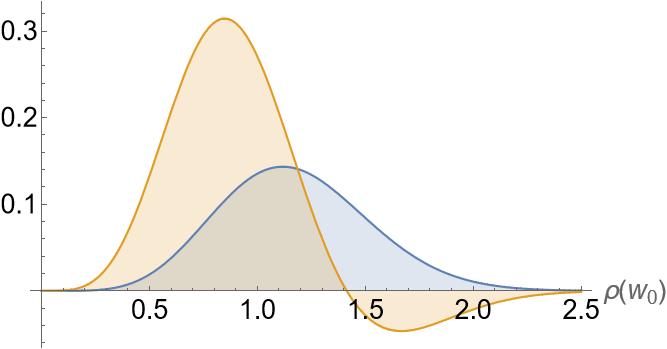}
\caption{Variations of the components of the angular momentum density ${\bar {\bf j}}$ (arbitrary units) with the radial coordinate $\rho$. The blue curve represents the variations of the transverse component (Eq. \ref{jayphi}) and the yellow curve represents the variations of the longitudinal component (Eq. \ref{jayzee}). For illustration we have chosen to consider the order $m=2$ and the vector mode for which $\Phi_p=0$ and $\cos\Theta_P=0.99$ of the order m=2 Poincar\'e sphere.}
\label{fig:taml2m3}
\end{figure}

 For illustration we now examine the distributions of the angular momentum density components for a representative case, namely the order $m=3$ Poincar\'e mode of the Laguerre-Gaussian type for which the mode function ${\tilde {\cal F}}$ is given by Eq.(\ref{laguerre}).  Figure (\ref{fig:saml2m3}) displays the variations of the SAM density components with the radial coordinate while Fig.(\ref{fig:taml2m3}) displays the corresponding angular momentum density variations with the radial coordinates.

\subsection{Integrated angular momentum density}
The space integral of the angular momentum density  over the x-y plane is formally defined by 
\beq
{\bn {\bar J}}=\int_0^{\infty}\rho\;d\rho\int_0^{2\pi}d\phi\left\{{{\bar j}}_{\bot}{\bn {\hat \phi}}+{\bar j}_z{\bn{\hat z}}\right\}={{\bar J}}_{\bot}{\bn {\hat \phi}}+{\bar J}_z{\bn{\hat z}}.
\eeq
The integral of the transverse density vanishes by virtue of the $\phi$ integral. We can then write
\beq
{{\bar J}}_{\bot}=0.
\eeq
The integral of the longitudinal term is 
\bea
{\bar J}_z&=&\frac{\epsilon_0 c^2 k_z^2}{2\omega}\nonumber\\
&\times&\int_0^{\infty}\rho d\rho\int_0^{2\pi}d\phi\left\{m{\tilde {\cal F}}^2+\rho{\tilde {\cal F}}{\tilde {\cal F}}'\right\}\cos\Theta_P.\label{Jzed}
\eea

The two different integrals involved in the evaluation of Eq.(\ref{Jzed}) are detailed in Appendix B.  The result is
\beq
{\bar J}_z=\frac{k_z w_0^2\pi{\cal E}_0^2}{4\mu_0 c}\left[\cos\Theta_P(m-1)\right],
\eeq
and on substituting for ${\cal E}_0$, we find
\bea
{\bar J}_z&=&\left(\frac{{\cal P}_T}{\omega c}\right)\left[\cos\Theta_P(m-1)\right].\label{Jz}
\label{Eqn_Jz}
\eea
The factor $\left(\frac{{\cal P}_T}{\omega c}\right)$ has the dimensions of angular momentum per unit length, as is the case with SAM. Note, however, that this total angular momentum depends on the order $m$. 
\textcolor{black}{The factor $(m-1)$ appearing in Eq.(\ref{Jz}) is consistent with the observation that in cylindrical coordinates the two terms in the polarisation vector in Eq.(\ref{epsil}) have phase functions $e^{i(m-1)\phi}$ and $e^{-i(m-1)\phi}$.} The factor $\cos(\Theta_P)$ takes the values $\pm 1$ at $\Theta_P=0,\pi$ as for circularly-polarised modes and is zero for $\Theta_P=\pi/2$, as for radially and azimuthally-polarised modes.  Other values of $\Theta_P$ concern elliptically-polarised modes. For $m=0$ and $\cos \Theta_P=1$ we recover the conventional result of ${\cal P}_T/(\omega c)$ appropriate for circularly polarised Gaussian modes.  For $m=1$, we recover the zero angular momentum  appropriate for pure radially and azimuthally-polarised vortex modes.

\subsection{Orbital angular momentum OAM}
\begin{figure}
\includegraphics[width=\columnwidth]{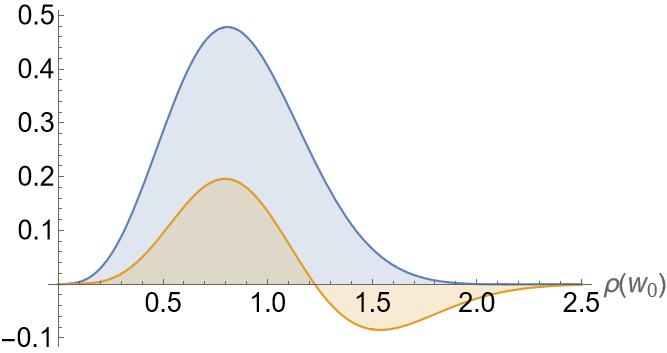}
\caption{Variations of the orbital angular momentum density (in arbitrary units) with the radial coordinate $\rho$. The blue curve represents the variations of the transverse component and the yellow curve the longitudinal component (Eq. \ref{jbar1}). For illustration we have chosen to consider the order $m=2$  and the vector mode close to  the north pole point for which $\cos\Theta_P=0.99$.}
\label{fig:oam}
\end{figure}

\textcolor{black}{ Having derived expressions for the AM density ${\bn {\bar j }}$ and the SAM density ${\bar {\bf s }}$, we consider the difference $({\bn {\bar j }}-{\bn {\bar s }})$ as representing OAM density ${\bar {\bf \ell }}$ \cite{allen1999,crimin2024}. 
 Thus in our case the OAM density components are  obtainable by subtracting the SAM density components, Eqs.(\ref{transverse}) and (\ref{longit}), from the angular momentum density components, Eqs.(\ref{jayphi}) and (\ref{jayzee}).  We have}
\begin{widetext}
\bea
\textcolor{black}{{\bn {\bar \ell}}={\bn {\bar j}}-{\bn {\bar s}}}&=&\textcolor{black}{\left(\frac{\epsilon_0 c^2 k_z}{2\omega}\right)\left\{k_z^2\rho {\tilde {\cal F}}^2-\frac{m{\tilde {\cal F}}^2}{\rho}-{\tilde {\cal F}'}{\tilde {\cal F}}\right\}{\bn {\hat{\phi}}}}\nonumber\\
&+&\textcolor{black}{\cos\Theta_P\left(\frac{k_z \epsilon_0c^2}{2\omega}\right)\left\{m {\tilde {\cal F}}^2+\rho{\tilde {\cal F}}{\tilde {\cal F}}'+{\tilde {\cal F}}^2\right\} {\bn {\hat{z}}}},
\label{jbar1}
\eea
\end{widetext}
and we note the two separate orbital angular momentum density components, namely the azimuthal component and the longitudinal one, both of which are seen to depend on the choice of  $m$ as well as the Poincar\'e angle $\Theta_P$. For illustration, we display in Fig.(\ref{fig:oam}) the variations of the OAM density components with the radial coordinate (in units of $w_0$).
\\
\\
\subsection{OAM per unit length}
The OAM per unit length is the space integral of the OAM density components over a cross section of the mode. Clearly the azimuthal component of the OAM component vanishes due to the angular integration.  The longitudinal component is then given by
\begin{widetext}
\beq
\textcolor{black}{{\bar L}_z=\cos\Theta_P\left(\frac{\pi k_z \epsilon_0c^2}{\omega}\right)\int_0^{\infty}\left\{m {\tilde {\cal F}}^2+\rho {\tilde {\cal F}}{\tilde {\cal F}}'+{\tilde {\cal F}}^2\right\}\rho\;d\rho.}
\label{ellbar1}
\eeq
\end{widetext}
For a Laguerre-Gaussian mode for which the amplitude function is given by Eq.(\ref{laguerre}) the integrals involved in the evaluation of Eq.(\ref{ellbar1}) are detailed in Appendix B.  It is easy to see that the integrals of the last two terms in Eq.(\ref{ellbar}) are ${\cal I}_Q$ and ${\cal I}_P$, which have the same magnitude, but opposite signs.  These terms now cancel identically and we obtain for the OAM density per unit length 
\beq
\textcolor{black}{{\bar L}_z=
m\cos\Theta_P\left(\frac{{\cal P}_T}{\omega c}\right).}
\label{Ellz}
\eeq
Thus for $m=0$ we have a Gaussian mode with zero OAM, so that the total angular momentum is entirely due to optical spin.  For $m\neq 0$ the OAM increases with increasing $m$, with $\cos\Theta_P$ spanning the range $(+1,-1)$ involving generally elliptical polarisations.  
\subsection{Optical helicity}

A recently published brief account of the helicity of higher order Poincar\'e modes appeared as a short communication by the current authors \cite{babiker24}. That communication provided little details of the procedure leading to the evaluations of the helicity density and  its space integral as the helicity per unit length. The formalism was also characterised by the use of cylindrical polar coordinates to describe the fields. Only the results for helicity density and helicity per unit length of the higher order modes for which $p=0$ were presented in \cite{babiker2024}.  However, we feel that it is necessary to discuss here the details of the evaluations specifically in Cartesian coordinates for the general case of $m\geq 0$ and $p\neq 0$ as the evaluations are both intricate and worthy of presentation using the Cartesian coordinate system we have adopted throughout this paper. As we shall see the results that emerge are more general and show that the inclusion of the radial number $p>0$ enhances the super-chirality, but we shall also show that we recover  the special case $p=0$ discussed in\cite{babiker24}.

The cycle-averaged optical helicity density and chirality density are defined generally in Eq.(\ref{hel}). 
There are thus contributions arising from the total electric field ${\bf E}=({\bf E}_1 +{\bf E}_2)$ and magnetic field ${\bf B}=({\bf B}_1 +{\bf B}_2)$, so that  
\bea
{\bar {\eta}}({\bf r})&=&\frac{c}{\omega^2}{\bar \chi}\label{chaid}\nonumber\\
&=&-\frac{\epsilon_0 c}{2\omega}\Im{[{\bf (E_1+E_2)}^*\cdot{\bf (B_1+B_2)}]},\label{equationa}
\eea
where ${\bf E}_{i}$ and ${\bf B}_{i}$, with $i=1,2$ are as given in Eqs.(\ref{emfields}). We focus on the helicity from which the chirality can be determined using  Eq.(\ref{chaid}) and proceed to evaluate both the helicity density and its space integral for a general higher order optical vortex mode.
The four terms arising from the expansion of Eq.(\ref{equationa}) are evaluated separately and now require as a first step expressions for the  x and y-derivatives of ${\cal F}^{(1)}$ and ${\cal F}^{(2)}$ in polar coordinates which are displayed in Eqs.(\ref{dfx}) to (\ref{dfy2}). 
It turns out that the sum  ${\bf E}_1^*\cdot{\bf B}_2+{\bf E}_2^*\cdot{\bf B}_1$ does not contribute an imaginary part and only the two direct terms ${\bf E}_1^*\cdot{\bf B}_1+{\bf E}_2^*\cdot{\bf B}_2$ contribute. 
 As an example how the evaluations proceed we consider the first dot product term.  We have using the first and second equations in Eq.(\ref{emfields})
\begin{widetext}
\bea
 [{\bf E_1}^*\cdot{\bf B_1}]
 &=&2ick_z^2|{\cal F}^{(1)}|^2+c\left\{i\left|\frac{\partial {\cal F}^{(1)}}{\partial x}\right|^2+i\left|\frac{\partial {\cal F}^{(1)}}{\partial y}\right|^2-\left(\frac{\partial {\cal F}^{(1)}}{\partial y}\right)^*\left(\frac{\partial {\cal F}^{(1)}}{\partial x}\right)+\left(\frac{\partial {\cal F}^{(1)}}{\partial x}\right)^*\left(\frac{\partial {\cal F}^{(1)}}{\partial y}\right)\right\}.
 \label{dot}
\eea
\end{widetext}
The next steps involve substituting for the  x- and y-derivatives of ${\cal F}^{(1)}$ using Eqs.(\ref{dfx}) and (\ref{dfy}).  This leads  
from  Eq.(\ref{dot}) to the result for the first term in the helicity density
\bea
\frac{1}{|{\cal U}_P|^2} \Im{[{\bf E_1}^*\cdot{\bf B_1}]}&=&2ck_z^2|{\tilde {\cal F}}|^2+c\left\{|{\tilde{\cal F}}'|^2+\frac{m^2}{\rho^2}|{\tilde {\cal F}}|^2\right\}\nonumber\\
 &+&c\frac{2m}{\rho}{\tilde{\cal F}}'{\tilde{\cal F}}.
 \label{direct1}
 \eea
 
 The second term $\Im{[{\bf E_2}^*\cdot{\bf B_2}]}$ follows the same steps to obtain for the dot product
 \bea
\frac{1}{|{\cal V}_P|^2} \Im{[{\bf E_2}^*\cdot{\bf B_2}]}&=&-\left(2ck_z^2|{\tilde {\cal F}}|^2+c\left\{|{\tilde{\cal F}}'|^2+\frac{m^2}{\rho^2}|{\tilde {\cal F}}|^2\right\}\right)\nonumber\\
 &-&c\frac{2m}{\rho}{\tilde{\cal F}}'{\tilde{\cal F}}.
 \label{direct2}
 \eea
 Thus we find
 \beq
 \frac{1}{|{\cal U}_P|^2} \Im{[{\bf E_1}^*\cdot{\bf B_1}]}=-\frac{1}{|{\cal V}_P|^2} \Im{[{\bf E_2}^*\cdot{\bf B_2}]}.
 \eeq

After some algebra, the results (\ref{direct1}) and (\ref{direct2}), together with the use of Eq.(\ref{poincare}), lead to the final expression for the helicity density 
\begin{widetext}
\beq
 {\bar {\eta}}({\bf r})=\frac{\epsilon_0 c^2}{4\omega}\cos{(\Theta_P)}\left\{2k_z^2|{\tilde{\cal F}}_{m,p}|^2+|{\tilde{\cal F}}'_{m,p}|^2
 +m^2\frac{|{\tilde{\cal F}}_{m,p}|^2}{\rho^2}+2m\frac{{\tilde{\cal F}}_{m,p}'{\tilde{\cal F}}_{m,p}}{\rho}\right\}.
 \label{heldens3}
\eeq
\end{widetext}
The first two terms of Eq.(\ref{heldens3}) are identifiable as the  zero order ($m=0$) helicity density for a general elliptical polarisation.  The rest of terms are the $m$-dependent higher-order terms.    

Recall that the Poincar\'{e} angle $\Theta_P$ spans the range $\Theta_P=0$ to $\Theta_P=\pi$ and within this range the Poincare function $\cos(\Theta_P)$ varies continuously from  +1.0 ($\Theta_P=0$, which corresponds to right-hand circular polarisation at the north pole), to -1.0; $\Theta_P=\pi$ (left-hand circular polarisation at the south pole).  Between the two pole points where $0<\Theta_P<\pi$ we have elliptical polarisation.  However when $\Theta_P=\pi/2$ and $m=1$, we have radial and azimuthal polarisation and we see that the helicity density vanishes at all points on the equatorial circle in this case and for all higher order $m$. 

  The general polarisation state covers a continuous set of points on longitudes on the surface of the order $m$ unit Poincar'e sphere.  In order to specify the helicity density we only need the order $m$, the angles $(\Theta_P, \Phi_P)$ and the optical vortex amplitude function ${\tilde {\cal F}}$. Note that the general helicity density Eq.(\ref{heldens3}) shows no dependence on $\Phi_P$. This means, for example, all points on a given latitude intersecting a given longitude point have the same helicity density.

A special case is that for which $m=0$ and $\Theta_P=0$ or $\pi$, for which all the $m$-dependent terms  of Eq.(\ref{heldens3}) are zero. The remaining expression corresponds to the helicity density of circularly-polarised general optical vortex mode \cite{babiker2022}.  This is because in this case the overall factor $\cos(\Theta_P)=\pm 1$ can be identified as $\sigma=\pm 1$, as for circular polarisation. 

 The helicity density at the general point ${\Theta_P,\Phi_P}$ as given by Eq.(\ref{heldens3}) and with $m\neq 0$ the $m$-dependent terms come into play for all values of $\cos{(\Theta_P)}$ in the range $(+1.0\; {\rm {to}}\; -1.0)$, corresponding to elliptically polarised modes (including circular,  linear, as well as radial and azimuthal).  In particular, 
for $m\geq 1$, as $\Theta_P$ increases from its value at the north pole, the function $\cos(\Theta_P)$ vanishes at $\Theta_P=\pi/2$  for all points $\Phi_P$ on the equatorial circle. It is only for the first order $m=1$ that this vanishing density corresponds to pure radially-polarised optical vortex modes. 

We may now evaluate the super-chirality properties of higher order for the special case involving  Laguerre-Gaussian modes.  The amplitude function for topological order $m$, radial number $p$ and waist $w_0$ is given by Eq.(\ref{laguerre}). The results are shown in Fig. 5 in which the helicity density variations are shown for different topological order $m$. Figure 5 displays the helicity density in this special case with topological order values $m=0,1,25,50, 100$ and $150$  for $\Theta_P=0$.  It is clear from Fig.5 that for $m=0$ and $m=1$ the helicity density is maximum on axis $\rho=0$ and that the case $m=1$ is more than double that for $m=0$, which makes the case $m=1$ super-chiral.  The $m=1$ helicity density variation contrasts with the cases $m\geq 2$ in which  the helicity density  vanishes at $\rho=0$ and the maxima occur off-axis. For $m\geq 2$ the density is initially marginally larger than the case $m=0$.  It is concentrated off-axis $\rho>0$ and it initially decreases with increasing $m$, but then increases with increasing $m$.

In Fig. 6 we focus on the helicity density variations of the lowest orders $m=0$ and $m=1$ and the $p$ values $p=0$ and $p=1$ using the same parameters as in Fig.5.  We see that the maximum of the first order $m=1,p=0$, as in Fig.5 exceeds that of $m=0,p=0$, However, the case $m=1;p=1$ shows a much higher value of the helicity density at $\rho=0$ as a main feature of this mode.  This observation confirms that this mode behaves like a Gaussian, as for $m=0$, and is  indicative of spin-to-orbit conversion such that $\sigma+m=0$.  It follows that since we have $m=1$, we must have  $\sigma=-1$ corresponding to one of the components forming the Poincar\'e polarisation ${\bn {\epsilon}}_{m=1,p}.$ \textcolor{black}{This is consistent with our earlier comment following Eq. (\ref{Eqn_Jz}) about the appearance of $(m-1)$ in the phase functions and the form of axial angular momentum component ${\bar J}_z$.}
\textcolor{black}{When we consider a larger $w_0$, as in Fig.7 where $w_0=\lambda$, we find that }\textcolor{black}{the on-axis helicity density  for the case $p=0$; $m=1$ is lower than that for the case $p=0$; $m=0$, in contrast with the corresponding case in Fig.6.  This is because the $p=0$; $m=1$ case is due to the longitudinal field components becoming weaker when the beam width is larger.}  \textcolor{black}{However, the enhanced on-axis helicity densities in the cases $p=1; m=0$ and $p=1; m=1$ are evidence of $p$-dependent and $m$-dependent contributions arising from the longitudinal components as radial and azimuthal gradients of the transverse fields, respectively.} 

\subsection{Integrated helicity}
We can evaluate the integral of  the helicity density  in  Eq.(\ref{heldens3})) over the $x-y$ plane.   
First we note that the radial integral of all terms in the form $\frac{{\tilde {\cal F}'}{{\tilde{\cal F}}}}{\rho}$ are identically zero for all mode functions which satisfy ${\tilde {\cal F}}_{\{m\}}(0)=0={\tilde {\cal F}}_{\{m\}}(\infty)$. The integrals needed to evaluate the helicity density per unit length are given in Appendix B for the special case of Laguerre-Gaussian mode for which amplitude function is given by Eq.(\ref{laguerre}). 
\begin{figure}
\includegraphics[width=1\linewidth]{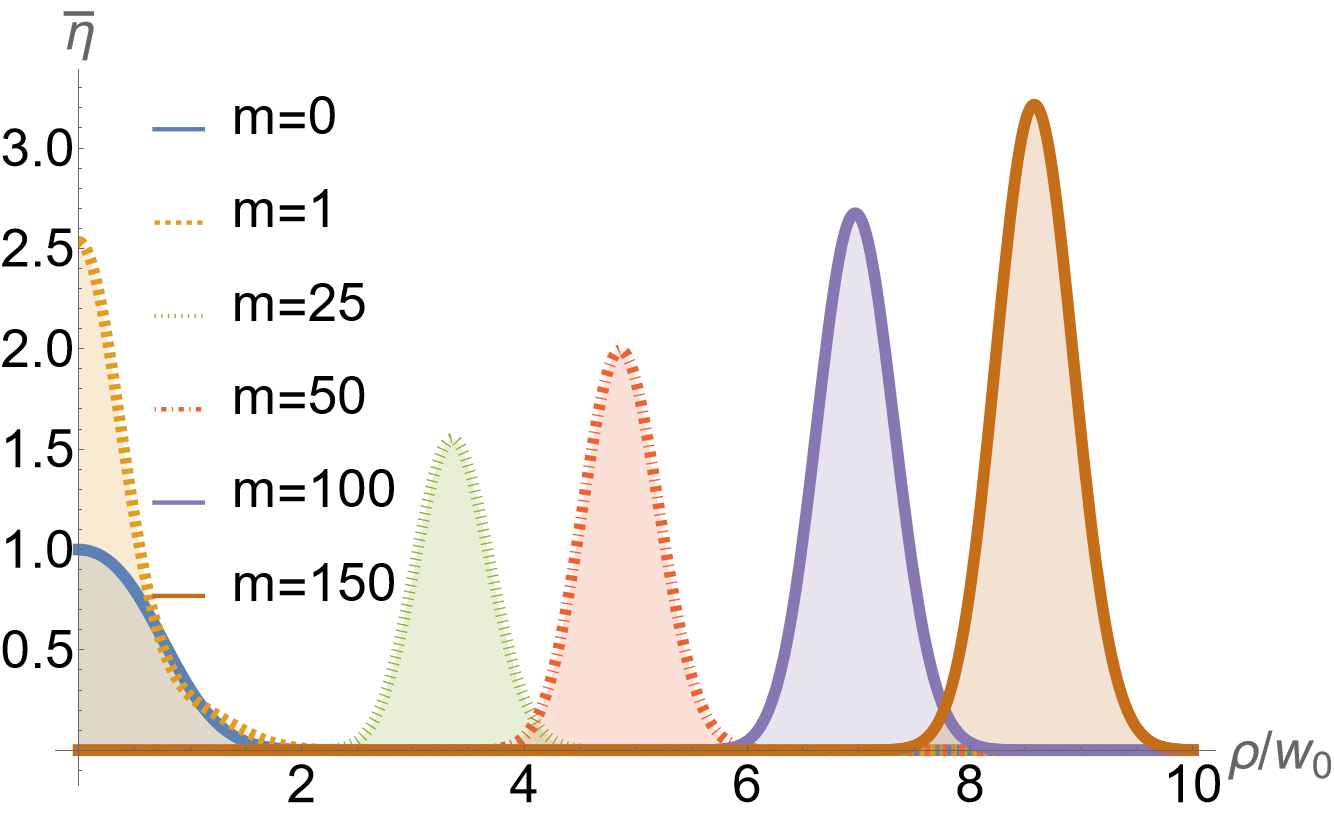}
\caption{Variations with the radial coordinate $\rho$ in units of $w_0$  of the helicity density, Eq.(\ref{heldens3}), due to focused Poincare modes of orders $m=0,1,25,50, 100$ and $150$ with $p=0$ and $w_0=0.5\lambda$.  Each of these modes lies on the longitude $(\Theta_P,\Phi_P=0)$, so that the values at a given point on this longitude should be multiplied by $\cos{(\Theta_P)}$. \textcolor{black}{Note that values on each curve are given relative to the maximum of the Gaussian helicity density ($m=0, p=0$; blue solid curve) which is set to the value of $1.0$. Superchirality for a given curve is said to occur when the helicity density at a given radial position exceeds $1.0$.}}
\end{figure}
\begin{figure}
\includegraphics[width=1\linewidth]{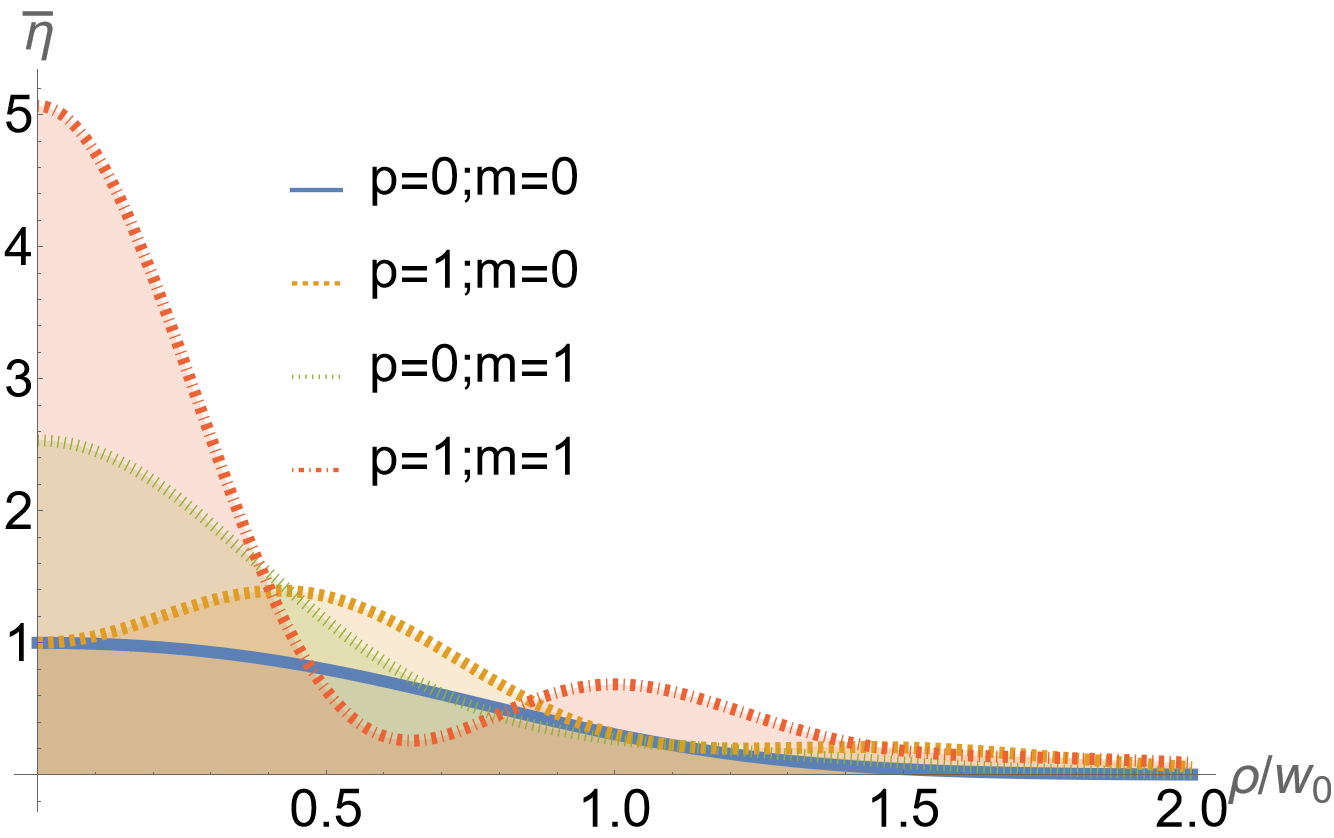}
\caption{Variations with the radial coordinate $\rho$ in units of $w_0$  of the helicity density, Eq.(\ref{heldens3}), due to focused Poincar\'e modes for which $m=0,1$ and $p=0,1$ with $w_0=0.5\lambda$. The modes lie on the longitude $(\Theta_P,\Phi_P=0)$, so that the values at any other given point on this longitude should be as shown, but multiplied by $\cos{(\Theta_P)}$.\textcolor{black}{Note that values on each curve are given relative to the maximum of the Gaussian helicity density ($m=0, p=0$; blue solid curve) which is set to the value of $1.0$. Superchirality for a given curve is said to occur when the helicity density at a given radial position exceeds $1.0$.} }
\end{figure}  

\begin{figure}
\includegraphics[width=1\linewidth]{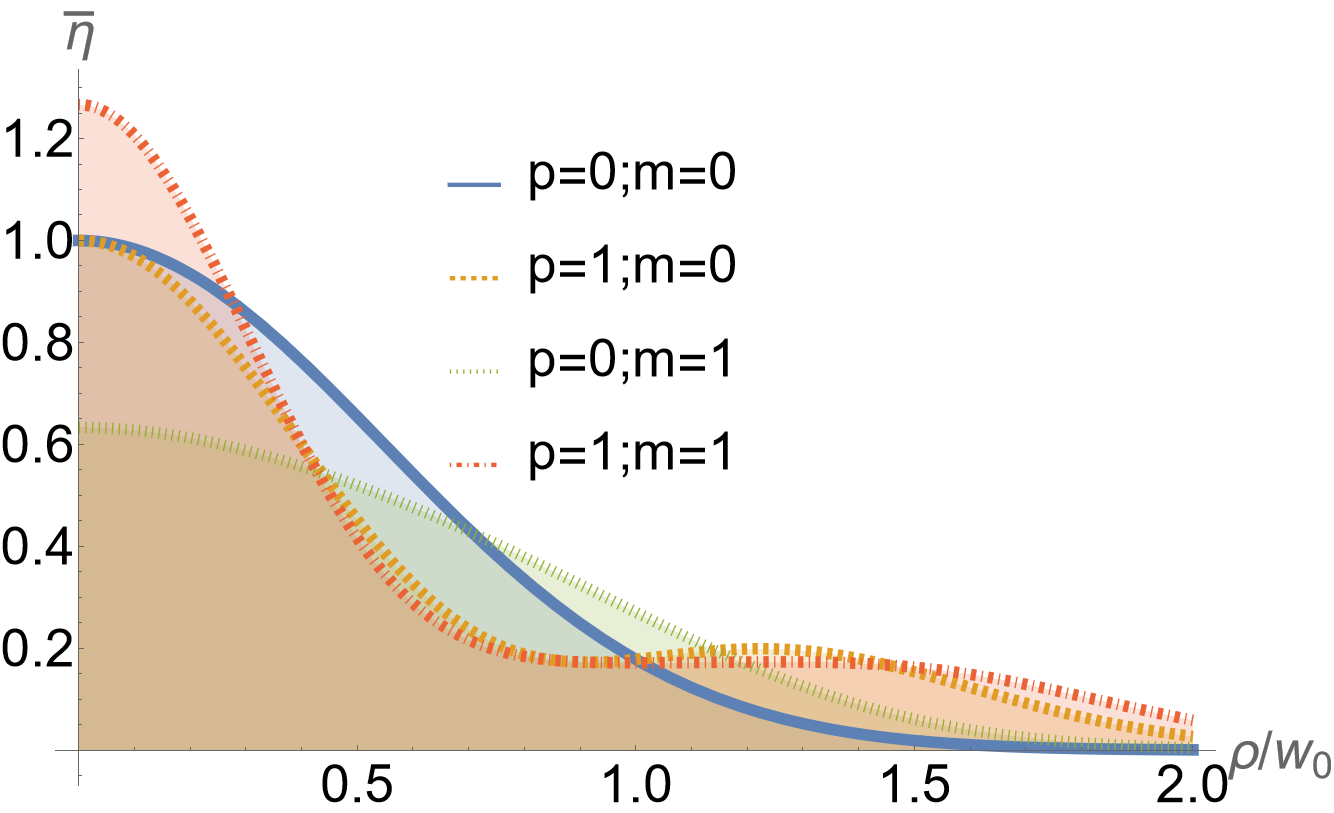}
\caption{As in Fig.6, but with a larger beam width  at focus $w_0=\lambda$. \textcolor{black}{Note the general reduction in all cases of the helicity density due to this. \textcolor{black}{But the peak helicity density for the case of $m=1$; $p=1$ still exceeds that of our reference case ($m=0$; $p=0$), even for this larger beam width.} See the main text for further remarks on this figure.} }
\end{figure}  

We are finally led to the helicity per unit length \textcolor{black}{at the focal plane } as
\beq
{\bar {\cal C}}_{m,p}
={\cal L}_0\cos{(\Theta_P)}\left(1+\frac{2 p+m+1}{k_z^2 w_0^2 }\right),
\label{final}
\eeq
where, as pointed out above,  ${\cal L}_0={\cal P}_T/(\omega c)$ is a constant for a fixed power ${\cal P}_T$ and we have substituted for ${\cal A}_0$ using Eq.(\ref{een0}). \textcolor{black}{The physical interpretation of the two terms within the brackets is that the first term accounts for the helicity due to the transverse fields which are formed as a superposition of two opposite circularly polarised vortex modes. The second term accounts for an additional helicity due to the longitudinal field components which arise from the non-zero gradient of the transverse field components in the same manner in which the Gouy phase arises (\cite{Feng2001}) as appears in Eq. (\ref{eq:refname4}) where we have the same $(2p+|m|+1)$ factor.}  Note that although the factor $1/k_z^2w_0^2$ in Eq.(\ref{final}) is typically small for $w_0^2\gg1/k_z^2$, the higher order helicity for which $m\gg 1$ and/or $p\geq 1$ would ensure super-chirality for relatively large $w_0$. 

\section{4. Summary}
The main mission of this paper has been to determine the intrinsic properties of Poincar\'e higher order modes. We have pointed out the significance of these modes for the processes of optical trapping of atoms and small particles and also in the manipulation of chiral matter and in providing improved encoding schemes for larger bandwidth optical quantum communications.  Here their intrinsic properties, as we emphasised, have previously neither been considered, nor their values determined. We have argued that our approach towards determining the intrinsic properties is most essential because it takes into account the participation of the axial components of the twisted light fields of the Poinar\'e modes.  \textcolor{black}{We have shown that  proper determinations of their angular momentum, both spin and orbital, their helicity and the chirality require the inclusion of the longitudinal field components}. We have predicted significant enhancements of the values of the intrinsic properties of these modes when compared with those due to the zero order optical modes.

Our first finding in this paper consists of determining the spatial distributions of the spin angular momentum (SAM) density components, both transverse and longitudinal of the higher order modes.  However on integrating the density components, we only find a non-zero longitudinal component which  is independent of $m$ .  This means that all higher order modes have the same  SAM which depends only on the Poincare angle $\Theta_P$. The factor $\cos(\Theta_P)$ takes the values $\pm 1$ at $\Theta_P=0,\pi$ as for circularly-polarised modes and is zero for $\Theta_P=\pi/2$, as for $m=1$ radially and azimuthally-polarised modes.  Other values of $\Theta_P$ concern elliptically-polarised modes.

Our second finding concerns the angular momentum (AM) density, which for paraxial light must be the sum of the SAM density and the OAM density.  We obtained the OAM density by subtracting the SAM density from the AM density.  We then evaluated the space integrals of both the AM and the OAM and found that they include new terms dependent on the order $m$ and on $\cos(\Theta_P)$.  However, these expressions reduce to the known zero order and first order results for any point on the surface of the respective order Poincar\'e sphere.

Finally, we tackled the helicity and chirality of these higher order vortex modes, obtaining a general expression for the helicity density and discussed helicity  for the relevant special lower order cases.  In particular, for points  $(\pi/2,\Phi_P)$ on the equatorial circle, the helicity (and so also chirality) is always zero, as for pure radially and azimuthally-polarised modes. The case $m=1,p=1$ displays the highest helicity density which is at least double the $m=0$ helicity density.  This  means that this mode exhibits super-chirality since it is enhanced relative to the helicity of an ordinary (order $m=0$) circularly-polarised mode. For all other points on the surface of the  Poincar\'{e} sphere the variations of the helicity density display an initial decrease with increasing $m$ followed by an increase.  
We have also shown that a higher order $m$ Laguerre-Gaussian mode for which $m=1,p=1$ is a strongly super-chiral vortex beam which is dominated by  the vortex core at $\rho=0$. We have found that other higher order Laguerre-Gaussian modes for which $m>1,p>0$ have off-axis maximum helicity  which is also super-chiral. These results strongly indicate the existence of a highly desirable super-chirality property of the higher order modes which, we suggest, is now ripe for direct experimental investigation.  There are diverse applications that can be envisaged, including improved interactions with chiral matter and stronger trapping and manipulation using optical spanners and tweezers, for example in micro-fluidics and improved encoding schemes for higher bandwidth quantum communications. 

\section*{Appendix A: Normalisation}

The higher order vortex mode normalisation is related to the applied power ${\cal P}_T$  evaluated as the space integral over the beam cross-section of the z-component of the Poynting vector.  We have 
\beq
{\cal P}_T=\frac{1}{2\mu_0}\int_0^{2\pi}d\phi\int_0^{\infty}|\left\{({\bf E}_1+{\bf E}_2)^*\times({\bf B}_1+{\bf B}_2)\right\}_z|\rho d\rho.
\label{eypeem}
\eeq
There are 4 terms, 2 direct and two mixed terms, but once again, we find that the mixed terms cancel. Consider first the mixed terms. We have
\bea
{\cal P}_{12z}&=&[{\bf E}_1^*\times{\bf B}_2]_z\nonumber\\
&=&E_{1x}^*B_{2y}-E_{1y}^*B_{2x}\nonumber\\
&=&{\cal U}_P^*{\cal V}_P\left\{(-ick_z)(ik)-(ck_z)(k_z)\right\}{\tilde {\cal F}}^2\nonumber\\
&=&0.
\eea
Similarly, we have 
\bea
{\cal P}_{21z}&=&[{\bf E}_2^*\times{\bf B}_1]_z\nonumber\\
&=&E_{2x}^*B_{1y}-E_{2y}^*B_{1x}\nonumber\\
&=&{\cal U}_P^*{\cal V}_P\left\{(-ick_z)(ik_z)-(-ck_z)(-k_z)\right\}{\tilde {\cal F}}^2\nonumber\\
&=&0.
\eea
Also the direct terms are as follows
\bea
{\cal P}_{11z}&=&[{\bf E}_1^*\times{\bf B}_1]_z\nonumber\\
&=&E_{1x}^*B_{1y}-E_{1y}^*B_{1x}\nonumber\\
&=&|{\cal U}_P|^2\left\{(-ick_z)(ik_z)+(ck_z)(k_z)\right\}{\tilde {\cal F}}^2\nonumber\\
&=&2ck_z^2{\tilde {\cal F}}^2|{\cal U}_P|^2,
\eea
and
\bea
{\cal P}_{22z}&=&[{\bf E}_2^*\times{\bf B}_2]_z\nonumber\\
&=&E_{1x}^*B_{1y}-E_{1y}^*B_{1x}\nonumber\\
&=&|{\cal V}_P|^2\left\{(-ick_z)(ik_z)+(ck_z)(k_z)\right\}{\tilde {\cal F}}^2\nonumber\\
&=&2ck_z^2{\tilde {\cal F}}^2|{\cal V}_P|^2.
\eea
Collecting terms, we find, using the last identity in Eq.(\ref{poincare})
\bea  
\left\{({\bf E}_1+{\bf E}_2)^*\times({\bf B}_1+{\bf B}_2)\right\}_z&=&2ck_z^2{\tilde {\cal F}}^2|(|{\cal U}_P|^2+|{\cal V}_P|^2)\nonumber\\
=ck_z^2{\tilde {\cal F}}^2.
\eea

Then Eq.(\ref{eypeem}) yields
\beq   
{\cal P}_T=\frac{\pi ck_z^2}{\mu_0}{\cal I}_P,
\eeq
where 
\beq 
{\cal I}_P=\int_0^{\infty}{\tilde {\cal F}}^2\rho\;d\rho,
\eeq
so we can write
\beq
{\cal I}_P=\frac{\mu_0{\cal P}_T}{\pi ck_z^2}\label{peeT}.
\eeq

This result applies to any optical vortex mode characterised by  an amplitude function ${\tilde{\cal F}}$.    The integral ${\cal I}_P$ is evaluated in Appendix B for the special case of a vector mode involving Laguerre-Gaussians and so we can now evaluate the normalisation constant ${\cal A}_0$ using Eqs.(\ref{peeT}) and (\ref{eyePx}).  We then have
\beq
{\cal A}_0^2=\frac{4\mu_0 {\cal P}_T}{\pi c k_z^2w_0^2}.\label{een0}
\eeq

\section*{Appendix B: Four Integrals}
There are four integrals that require evaluation for the special case of a Laguerre-Gaussian mode for which ${\tilde{\cal F}}$ is given by Eq.(\ref{laguerre}).  These are 
\bea
&{\cal I}_P&=\int_0^{\infty}{\tilde {\cal F}}^2\rho\;d\rho\nonumber,\\
&{\cal I}_Q&=\int_0^{\infty}\rho^2{\tilde {\cal F}'}{\tilde {\cal F}}\;d\rho\nonumber,\\
&{\cal I}_R&=\int_{0}^{\infty}\frac{1}{\rho}{\tilde {\cal F}}^2 d\rho\nonumber,\\
&{\cal I}_S&=\int_{0}^{\infty}|{\tilde {\cal F}'}|^2\rho\; d\rho.
\eea
Substituting for ${\tilde{\cal F}}$ from Eq.(\ref{laguerre}) and using the variable $x=2\rho^2/w_0^2$ we have for ${\cal I}_P$
\beq
{\cal I}_P={\cal A}_0^2\frac{w_0^2}{4}\frac{(p+|m|)!}{p!}\int_0^{\infty}e^{-x}x^{|m|}[L^{|m|}_p(x)]^2 dx=\frac{{\cal A}_0^2w_0^2}{4}\label{eyePx},
\eeq
where the x-integral is a standard integral of the associated Laguerre-functions.

Consider next the evaluation of ${\cal I}_Q$.  We have on substituting for ${\tilde{\cal F}}$ from Eq.(\ref{laguerre}) and using the variable $x=2\rho^2/w_0^2$ 
\beq
{\cal I}_Q=\frac{w_0^2}{2}\int_0^{\infty}x{\tilde{\cal F}}(x){\tilde{\cal F}'}(x)dx,
\label{eyeqx}
\eeq
where
\beq
{\tilde {\cal F}}(x)={\cal A}_0^2\sqrt{\frac{p!}{(p+|m|)!}}x^{|m/2|}e^{-x/2}L^{|m|}_p(x).
\label{effofx}
\eeq
Integrating by parts in Eq.(\ref{eyeqx}) we have 
\bea
{\cal I}_Q&=&\frac{w_0^2}{2}\int_0^{\infty}\frac{x}{2}\left(\frac{d{\tilde {\cal F}}^2(x)}{dx}\right)dx\nonumber\\
&=&\frac{w_0^2}{4}\left\{x{\tilde{\cal F}}(x)^2\right\}_0^{\infty}-\frac{w_0^2}{4}\int_0^{\infty}{\tilde{\cal F}}^2(x)dx.
\eea
The first term yields zero at both limits and ${\cal I}_Q$ is given by  the second term.  On substituting for ${\tilde{\cal F}}(x)$ using Eq.(\ref{effofx}), we have
\bea
{\cal I}_Q&=&-\frac{w_0^2}{4}{\cal A}_0^2\frac{p!}{(p+|m|)!}\int_0^{\infty}x^{|m|}e^{-x}[L^{|m|}_p(x)]^2dx\nonumber\\
&=&-\frac{w_0^2}{4}{\cal A}_0^2=-{\cal I}_P.
\eea

Next we consider ${\cal I}_R$.  Substituting for ${\tilde{\cal F}}$ from Eq.(\ref{laguerre}) and using the variable $x=2\rho^2/w_0^2$ we have for ${\cal I}_R$
\bea
{\cal I}_R&=&\int_{0}^{\infty}\frac{1}{\rho}{\tilde {\cal F}}^2 d\rho\nonumber\\
&=&\frac{p!}{2(p+|m|)!}\int_0^{\infty} x^{|m|-1}e^{-x}[L^{|m|}_p(x)]^2dx\nonumber\\
&=&{\cal A}_0^2\frac{1}{2|m|}\label{ceebar}.
\eea
Finally we consider ${\cal I}_S$.  On substituting for the first derivative ${\tilde {\cal F}}'$, this integral splits into a number of terms which are then evaluated separately.  There are cancellations between the integrals of those terms and the result turns out to be
\bea
{\cal I}_S&=&\int_{0}^{\infty}{\tilde {\cal F}}'^2 \rho d\rho=\frac{2 p+1}{2} {\cal A}_0^2 \label{ceebar}.
\eea
\bibliography{References.bib}
\end{document}